\documentclass{optica-article}

\journal{opticajournal} 

\articletype{Research Article}

\usepackage{lineno}
\usepackage{braket}

\begin{document}

\title{Topological Edge and Corner States in Biphenylene Photonic Crystal}

\author{Huyen Thanh Phan,\authormark{1,*} Keiki Koizumi,\authormark{1} Feng Liu,\authormark{2,3,$\dagger$} and Katsunori Wakabayashi\authormark{1,4,5,$\ddagger$}}

\address{\authormark{1}Department of Nanotechnology for Sustainable Energy, School of Science and Technology, Kwansei Gakuin University, Gakuen-Uegahara 1, Sanda, Hyogo 669-1330, Japan\\
\authormark{2}School of Physical Science and Technology, Ningbo University, Ningbo 315-211, China\\
\authormark{3}Institute of High Pressure Physics, Ningbo University, Ningbo 315-211, China \\
\authormark{4}Center for Spintronics Research Network (CSRN), Osaka University, Toyonaka 560-8531, Japan\\
\authormark{5}National Institute for Materials Science (NIMS), Tsukuba, Ibaraki 305-0044, Japan}

\email{\authormark{*}phanthanhhuyenpth@gmail.com}
\email{\authormark{$\dagger$}liufeng@nbu.edu.cn}
\email{\authormark{$\ddagger$}waka@kwansei.ac.jp}


\begin{abstract*} 
The biphenylene network (BPN) has a unique two-dimensional atomic
 structure, where hexagonal unit cells are arranged on a square
 lattice. Inspired by such a BPN structure, we design a counterpart in
 the fashion of photonic crystals (PhCs), which we refer to as the BPN PhC. We
 study the photonic band structure using the finite 
 element method and characterize the topological properties of the BPN PhC
 through the use of the Wilson loop. 
Our findings reveal the emergence of topological edge states in the BPN
 PhC, specifically in the zigzag edge and the chiral edge, as a
 consequence of the nontrivial Zak phase in the corresponding
 directions. In addition, we find the
 localization of electromagnetic waves at the corners formed by the
 chiral edges, which can be considered as second-order topological
 states, i.e., topological corner states. 
\end{abstract*}

\section{Introduction} \label{sec1}
Topology originates from mathematics, which helps distinguish the
geometric structure by invariant quantities during continuous
deformation. For example, a closed surface is classified by the number
of hole in it. This number remains unchanged without cutting the
surface. When topology is applied to
physics~\cite{Bansil2016,Hasan2010,Qi2011,Ando2013}, some novel
properties are found to be robust to perturbations and have potential
applications in the realization of disorder-free spin
transport~\cite{Sato2016, Kong2010, Bernevig2006, Kim2019, Fu2007,
Kane2005, Chen2009, Okuyama2019}. One of the most important features on
nontrivial topological systems is the emergence of edge
states~\cite{Delplace2011,Obana2019, Yoshida2019, Song2020a,
Waka2009}. These topologically protected edge states are characterized
by the nontrivial energy band inversions at high symmetric points and
become a strong candidate for further study of quantum
computation. Moreover, recent studies of higher-order topology have
extended these topological edge states to the corner states in
two-dimensional (2D)
systems~\cite{Song2017,Benalcazar2017,Liu2021,Miao2022}. This research
direction becomes increasingly attractive because of its potential
applications. For example, quantum computation can be designed based on
topological corner
states~\cite{Harari2018,Wu2019,Zhang2020,Wu2020,He2019}. 

By applying topology to photonic crystalline systems, the photonic
analogy of the quantum Hall effect was theoretically proposed by Raghu
and Haldane~\cite{Haldane2008, Raghu2008}, then experimentally observed
in 2D photonic crystals (PhCs)~\cite{Wang2008,Wang2009}. These seminal works have stimulated
many other investigations on the topological edge states of
PhCs~\cite{Hafezi2011,Khanikaev2013,Hafezi2013,Rechtsman2013,Skirlo2014}.  
After that, researchers have become more and more interested in the
studies of topological PhCs. The photonic analogy of Chern insulators is
also found due to non-zero Berry curvature in broken time-reversal
symmetry~\cite{Wang2008,He2014,Blanco2020}. In addition, even in the
zero Berry curvature, topological states can emerge in the PhC
systems~\cite{Liu2018,Chen2019,Ota2019}. These states are explained by
nontrivial Zak phase~\cite{He2022,Chen2018}, gapped Wannier
bands~\cite{Liu2019,Xue2019,Wang2021, Zhang2021,Shao2023} and valleys interaction~\cite{Phan2021}. 

Designing a topologically nontrivial PhC usually requires the breaking
of discrete particle symmetry, such as the time-reversal symmetry, which
is not easy to achieve. Another routine is to mimic the existing
structures in electronic materials such as
graphene~\cite{Kane2005,Waka2009,Yuji2011}. 
Following the successful applications of graphene and other
nanoelectronic materials, the search for 2D carbon allotropes other than
graphene, such as biphenylene network (BPN) or graphenylene network, has stimulated and
brought a new insight into electronic transport properties in nanoscale
materials. 
Especially, BPN has a fascinating lattice structure, where the hexagonal unit cells are
organized on a square lattice, 
resulting in a two-dimensional tiling pattern that includes four-, six-, and eight-membered rings~\cite{Fan2021,Bafekry2021}.
Motivated by recent experimental development of the
BPN sheet, we design a PhC that has an atomic
structure and arrangement similar to those of the BPN and focus on its
topological properties such as the emergence of topological edge and
corner states. 
In this work, we use the finite element method to evaluate the photonic
band structures of BPN PhC
and utilize the Wilson loop to analyze their topological properties.
We find several different ribbon structures induce the topological edge
states which are attributed to the the presence of a nontrivial Zak phase.
Furthermore, when the product of the Zak phases for two
edge structures forming a corner is non-zero, it results in the
appearance of second-order topological states~\cite{Ota2019, Jin2022,Xie2018}. These results suggest
potential applications of topological edge and corner states in PhC
devices.

The paper is organized as follows. In Section 2, we introduce BPN PhC
structure and numerically calculate the photonic band structure. Section
3 is used to evaluate the Zak phase and discuss the existence of
topological edge states. The second-order topology as corner states is
presented in Section 4. We discuss the results and summarize the paper
in Section 5.

\section{Biphenylene Photonic Crystal}
\begin{figure}[ht!]
\centering\includegraphics[width=1.0\textwidth]{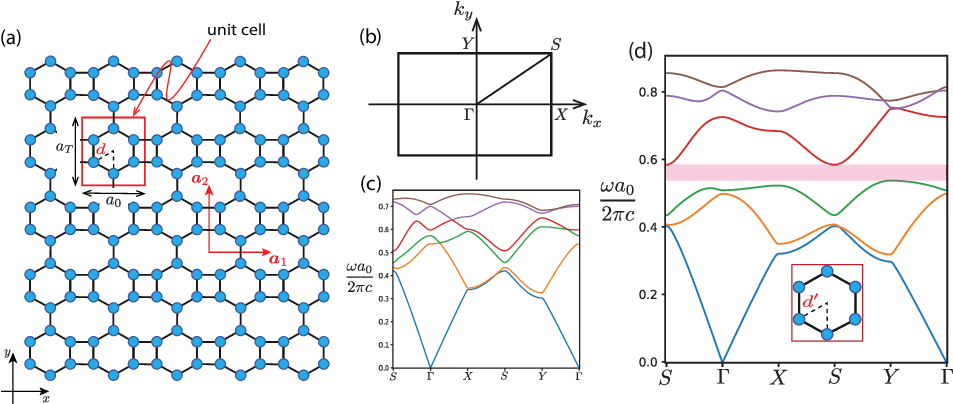}
\caption{(a) Schematic of 2D BPN PhC. 
 The red rectangle indicates the primitive unit cell, which contains six
 equivalent dielectric rods, colored by cyan. 
The radius of these rods is $r_0$, and the distance between each rod and
 the center of the unit cell is $d$. 
$\mathbf{ a_1}=(a_0,0)$ and
 $\mathbf{ a_2}= (0,a_T)$  are primitive lattice vectors.
Here, $a_0$ and $a_T\equiv {3a_0}/\left(\sqrt{3}+1\right)$ are the lattice
 periodicity in the $x$- and $y$-directions, respectively.  
(b) 1st BZ of the 2D BPN lattice. (c) 
Photonic band structure of 2D BPN PhC for $d={a_T}/3$,
 $r_0=0.2d$. (d) Photonic band structure for the modified BPN PhC 
where $d^\prime=1.3a_T/3$, $r_0=0.2d$. The
 inset shows the schematic of one unit cell of modified BPN PhC.} 
\label{figure1}
\end{figure}

Here we introduce BPN PhC and calculate its photonic band
structure. Figure~\ref{figure1}(a) shows the lattice structure of BPN
PhC, where the red rectangle indicates the primitive unit cell. 
There are six equivalent dielectric rods in one unit cell, colored by
 cyan. Radius for each rod is $r_0$. In this paper, we assume that the
 dielectric rods are all made of silicon, with a dielectric constant of
 $\varepsilon = 11.7$. These rods are arranged periodically in form of
 the BPN
 lattice, where the lattice periodicity in $x$- and $y$-directions are $a_0$ and
 $a_T={3a_0}/(\sqrt{3}+1)$, respectively. The
 primitive vectors of BPN PhC are $\mathbf{a}_1=(a_0,0)$ and 
 $\mathbf{a}_2= (0,a_T)$.
The distance between the two nearest dielectric rods is $d$, which is
 equal to the distance between each rod and the center of the unit cell.  
We focus on the transverse magnetic modes of the electromagnetic (EM) waves so that the
 magnetic field is in the $xy$ plane, the electric field is
 perpendicular to the $xy$ plane. Assuming the harmonic oscillation in 
 time dependence, the eigenvalue equation of the PhC is given as 
\begin{equation}
    \frac{1}{\varepsilon \left(\mathbf{ r} \right)}
     \,\mathbf{ \nabla} \times \mathbf{ \nabla} \times
     \mathbf{ E}\left(\mathbf{ r}\right) =
     \frac{\omega^2}{c^2}\mathbf{ E}\left(\mathbf{ r}\right),
\end{equation}
where $\mathbf{ E}\left(\mathbf{ r}\right)$ is the electric
field and $\varepsilon \left(\mathbf{ r} \right)$ is dielectric
function. $\mathbf{ r}=(x,y)$ is the position vector in 2D space.
$c$ is the speed of light in vacuum. $\omega$ is the eigenfrequency.
Figure~\ref{figure1}(b) shows the corresponding 1st Brillouin zone (BZ) for 2D
BPN lattice with high symmetric points.

The photonic band structure for
$d=a_T/3$ and $r_0=0.2d$ are shown in
Fig.~\ref{figure1}(c). As can be seen, there is no band gap in this PhC
structure, all bands connect to others by linear dispersion at
degenerate points. To examine the topological properties of BPN PhC, we
need to open a complete band gap in the photonic band
structure. Therefore, we slightly modify the lattice structure by
increasing the distance $d$ between each rod and the center of the unit
cell, i.e. $d^{\prime} = 1.3d$. The radii of all rods and the 
size of the unit cell remain the same as the original structure, which
can be seen in the inset of Fig.~\ref{figure1}(d). 

The modified photonic band structure is shown in
Fig.~\ref{figure1}(d). A complete band gap, labeled by shaded pink
color, is found around the normalized frequency $0.55$ and is between
the 3rd and the 4th band. 
The 1st and the 2nd photonic bands are still connected by a degenerate point
along the $S-Y$ line. The 3rd band is now isolated from the others with
small photonic band gap. For simplicity, we will examine the modified
BPN PhC and focus on the three lowest photonic bands below the band gap. 

\section{Zak Phase and Topological Edge States} \label{sec3}
In this section, we elaborate the numerical calculation method used
to evaluate the Zak phase. Then, we discuss
the localization of EM waves at the edges 
for several ribbon structures. 

The Zak phase is commonly calculated for one-dimensional (1D) periodic
systems. It is mathematically defined as the 1D integration
of the Berry connection over the 1st BZ~\cite{Zak1989}.
Since this integration occurs along a single dimension, 
the Zak phase of 2D periodic systems depends
on the other direction that is perpendicular to the integration
line~\cite{Delplace2011}, which can also be considered as a Wilson
loop. For instance, within the 1st BZ of a 2D square lattice, the
Zak phase along the $k_x$ direction should be determined by integrating
the Berry connection along the $k_y$ direction. 

\begin{figure}[ht]
\centering\includegraphics[width=0.8\textwidth]{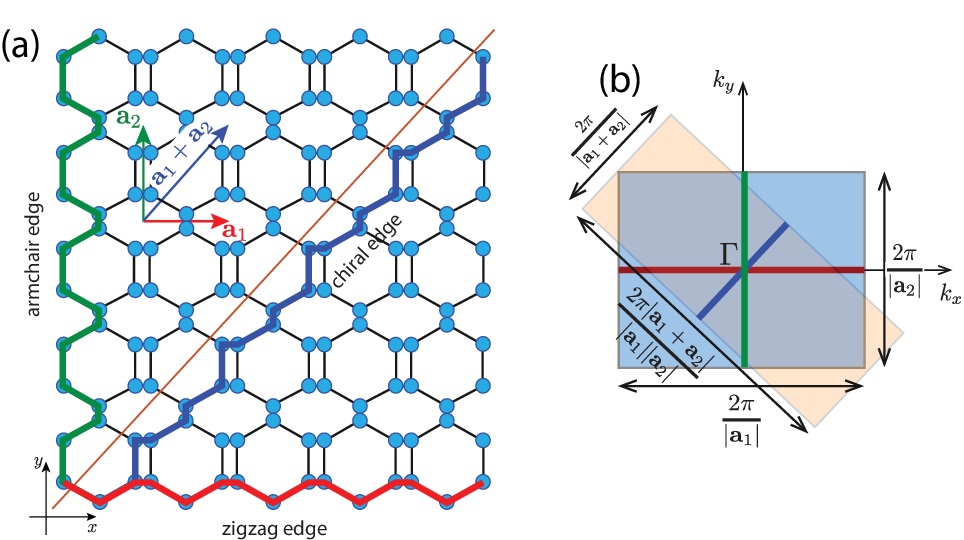}
\caption{(a) Schematic of edge structures for BPN PhC. The red
 line indicates the zigzag edge, whose translational vector is $\mathbf{
 T}=\mathbf{ a}_1$. The armchair edge is expressed as the 
 green line with translational vector $\mathbf{ T}=\mathbf{ a}_2$. The blue
 line denotes the chiral edge, whose translational vector is $\mathbf{ T}=\mathbf{ a}_1+\mathbf{ a}_2$. 
(b) The momentum areas used for calculation of Zak phase of BPN PhC with edges.
The blue rectangle is momentum area used for the calculation of Zak phase for the ribbons with the zigzag and armchair
edges. The orange rectangle is the momentum area used for the
 calculation of Zak phase for the ribbons with chiral edges.
 The size of each momentum area is written in black text.
The red, green and blue bold lines are
 the 1st BZs of ribbons with the zigzag, armchair and chiral edges,
 respectively.}  
\label{figure2}
\end{figure}

To evaluate numerically the Zak phase of BPN edges with several
different edge structures, we shall briefly illustrate the lattice
structure of BPN ribbons and their relevant structural parameters.
In Fig.~\ref{figure2}(a), we present three typical edge structures of BPN
ribbons, which are referred as zigzag, armchair and chiral edges.
These are indicated by red, green and blue lines, respectively.
For zigzag edges, the lattice is translational invariant under the
translational vector $\mathbf{T}=\mathbf{a}_1$. 
Similarly, for armchair edges, the translational vector is 
$\mathbf{T}=\mathbf{a}_2$, and for chiral edges,
$\mathbf{T}= \mathbf{a}_1 + \mathbf{a}_2$.

In general, we assume that the translational vector of BPN edges can
be defined as 
$\mathbf{T} = m \mathbf{a}_1 + n \mathbf{a}_2$,
where $\mathbf{a}_1$ and $\mathbf{a}_2$ are two primitive lattice vectors, $m$
and $n$ are coprime integers. If they are not coprime, the two vectors
defined below do not form the basis set in reciprocal space.  
Normally, the 2D momentum area that is used to calculate the Zak phase
is the 1st BZ. In more a general case, that 2D momentum area is chosen
as a rectangular area formed by two orthogonal vectors
$\mathbf{\Gamma_{\parallel}}$ and $\mathbf{\Gamma_{\perp}}$. These two vectors are
defined as follows 
\begin{equation}
    \mathbf{\Gamma_{\parallel}} = 2\pi\frac{ \mathbf{T}}{|\mathbf{T}|^2}.
\end{equation}
$\mathbf{\Gamma_{\parallel}}$ is parallel to the edge orientation
 (parallel to $\mathbf{ T}$).
Thus, the 1st BZ for BPN edges can be defined as
 $\displaystyle{-\frac{\pi}{|\mathbf{T}|}\le k\le \frac{\pi}{|\mathbf{T}|}}$.
The 2nd vector $\mathbf{ \Gamma_{\perp}}$ is perpendicular to $\mathbf{ \Gamma_{\parallel}}$ and relates to $\mathbf{ \Gamma_{\parallel}}$ by the following formula
\begin{equation}
\label{area}
    \mathbf{ \Gamma_{\parallel}} \times \mathbf{ \Gamma_{\perp}} = \mathbf{ b}_1 \times \mathbf{ b}_2,
\end{equation}
where
$\mathbf{ b}_1$ and $\mathbf{ b}_2$ are two primitive vectors in reciprocal
space and defined by $\mathbf{ a}_i \cdot \mathbf{ b}_j = 2\pi \delta_{ij}$
($i,j =1,2$). Equation~(\ref{area}) indicates that the area of the new
momentum area formed by $\mathbf{ \Gamma_{\parallel}}$ and $\mathbf{
\Gamma_{\perp}}$ is equal to the area of the original 1st BZ.
All the translational and reciprocal vectors for zigzag, armchair and
chiral edges are summarized in Table~\ref{tab.edge}. 

\begin{table}[htbp]
\centering
\caption {\bf Translational and reciprocal vectors for zigzag, armchair
 and chiral edges.}
\begin{tabular}{cccc}
\hline
Edge Structure & $\mathbf{T}$ & $\mathbf{\Gamma}_\parallel$ & $\mathbf{\Gamma}_\perp$ \\
\hline
Zigzag & $\mathbf{a}_1$ &
	 $\left(\displaystyle{\frac{2\pi}{a_0}},0\right)$ & $\left(0,\displaystyle{\frac{2\pi}{a_T}}\right)$\\
Armchair & $\mathbf{a}_1$  &
	 $\left(0,\displaystyle{\frac{2\pi}{a_T}}\right)$ & $\left(\displaystyle{\frac{2\pi}{a_0}},0\right)$\\
Chiral & $\mathbf{a}_1+\mathbf{a}_2$ & $\displaystyle{\frac{2\pi}{a_0}\left(\frac{4+2\sqrt{3}}{13+2\sqrt{3}}, \frac{3+3\sqrt{3}}{13+2\sqrt{3}}\right)}$ & $ \displaystyle{\frac{2\pi}{a_0}\left(1, -\frac{1+\sqrt{3}}{3}\right)}$\\
\hline
\end{tabular}
  \label{tab.edge}
\end{table}

In Fig.~\ref{figure2}(b), we display the momentum areas, which
are used to calculate the Zak phase. The blue rectangle is the momentum
area used for calculating the Zak phase of the zigzag and armchair
edges, which is completely overlapped with the 1st BZ of bulk 2D BPN. The red and
green bold lines are the 1st BZ for zigzag and armchair edges,
respectively. The orange rectangle is the momentum area for computing
the Zak phase of the chiral edge, the blue bold line denotes the
1st BZ for chiral edges.  
The size of each momentum area is expressed as the black text in the figure.

The electric field propagating in the PhCs can be written in term of Bloch wave function as $\mathbf{ E}^n_\mathbf{ k}\left(\mathbf{ r}\right) = \mathbf{u}^n_\mathbf{ k}\left(\mathbf{ r}\right) e^{i \mathbf{ k}\mathbf{ r}}$, where $\mathbf{u}^n_\mathbf{ k}\left(\mathbf{ r}\right)$ is the periodic function with the same periodicity as the PhCs.
Before calculating the Zak phase, we define the scalar product of two periodic functions as 
\begin{equation} 
\label{innerproduct}
     \braket{\mathbf{u}^n_{\mathbf{ k}}|\mathbf{u}^n_{\mathbf{ k^\prime}}} 
     = \int_{\text{unit cell}} {\mathbf{u}^n_{\mathbf{ k}}\left(\mathbf{ r}\right)}^{\ast}  \mathbf{u}^n_{\mathbf{ k^\prime}}\left(\mathbf{ r}\right) d\mathbf{r},
\end{equation}

For the edge in any direction characterized by vector $\mathbf{ T}$, the Zak phase of the $n$-th band is computed by integrating the Berry connection over $\mathbf{ \Gamma_{\perp}}$ as
\begin{equation} 
\label{zak}
    Z^n\left(k_{\parallel}\right) = \oint_{\mathbf{ \Gamma_{\perp}}} \braket{\mathbf{u}^n_{\mathbf{ k}} |i\partial_{\mathbf{ k}}|\mathbf{u}^n_{\mathbf{ k}}} dk_{\perp},
\end{equation}
where 
$k_{\parallel}$ and $k_{\perp}$ are $k$-points in $\mathbf{ \Gamma_{\parallel}}$ and $\mathbf{ \Gamma_{\perp}}$ directions, respectively.
For numerical calculation, we make an approximation for Eq.~(\ref{zak}) by dividing  $\mathbf{ \Gamma_{\parallel}}$ and $\mathbf{ \Gamma_{\perp}}$ into $N_0$ segments then taking the sum of the contribution of each segment. We obtain the discrete formula similar to the Wilson loop as shown below
\begin{equation} 
\label{zak2}
    Z^n\left(k_{i}\right) = -\operatorname{Im} \left( \log \prod_{k_{j}} \braket{\mathbf{u}^n_{k_{i},k_{j}}
    |\mathbf{u}^n_{k_{i},k_{j+1}}} \right),
\end{equation}
where $k_i$ is discrete $k$-point of $\mathbf{ \Gamma_{\parallel}}$ and $k_j$ is discrete $k$-point of $\mathbf{ \Gamma_{\perp}}$, ($i,j=1,...,N_0$). The Eq.~(\ref{zak2}) indicates that the Zak phase can be calculated for each $k_i$ point in the $\mathbf{ \Gamma_{\parallel}}$ direction. Equation~(\ref{zak2}) is used for a single band. For a group of degenerate bands, the scalar products are replaced by overlap matrices~\cite{Blanco2020, Wang_2019}. 
The overlap matrix $S$ for the group of $N$ degenerate bands between two $k$-point $\mathbf{ k}_1$ and $\mathbf{ k}_2$ is 
\begin{equation} 
\label{omatrix}
S_{\mathbf{ k}_1 \mathbf{ k}_2} =
\begin{bmatrix}
\braket{\mathbf{u}^1_{\mathbf{ k}_1} |\mathbf{u}^1_{\mathbf{ k}_{2}} } & \braket{\mathbf{u}^1_{\mathbf{ k}_{1}} |\mathbf{u}^2_{\mathbf{ k}_{2}} } & ... &\braket{\mathbf{u}^1_{\mathbf{ k}_{1}} |\mathbf{u}^{N}_{\mathbf{ k}_{2}} } \\
\braket{\mathbf{u}^2_{\mathbf{ k}_{1}} |\mathbf{u}^1_{\mathbf{ k}_{2}} } & \braket{\mathbf{u}^2_{\mathbf{ k}_{1}} |\mathbf{u}^2_{\mathbf{ k}_{2}} } & ... &\braket{\mathbf{u}^2_{\mathbf{ k}_{1}} |\mathbf{u}^{N}_{\mathbf{ k}_{2}} }\\
... & ... & ... & ...\\
\braket{\mathbf{u}^{N}_{\mathbf{ k}_{1}} |\mathbf{u}^1_{\mathbf{ k}_{2}} } & \braket{\mathbf{u}^{N}_{\mathbf{ k}_{1}} |\mathbf{u}^2_{\mathbf{ k}_{2}} } & ... &\braket{\mathbf{u}^{N}_{\mathbf{ k}_{1}} |\mathbf{u}^{N}_{\mathbf{ k}_{2}} }
\end{bmatrix},
\end{equation}
where the index $l$ of $\mathbf{u}^{l}_{\mathbf{ k}}$ indicates band index.


Take the summation product over the period along the $k_j$, we obtain the Wilson-loop-like matrix as
\begin{equation} 
\label{zak3}
    \hat{S}\left(k_{i}\right) = \prod_{k_{j}} S_{k_i k_j , k_i k_{j+1}},
\end{equation}
The Zak phase for $n$-th band can be then solved by the $n$-th eigenvalues $s^n$ of the Wilson loop matrix $\hat{S}\left(k_{i}\right)$. The Zak phase for $n$-th subband is given by 
\begin{equation} 
\label{zak4}
    Z^n \left(k_{i}\right) = -\operatorname{Im} \log
    \left(s^n\right).
\end{equation}
As can be seen in photonic band structure in Fig.~\ref{figure1}(d), the
1st and the 2nd photonic bands are entangled with a degenerate
point, however, the 3rd photonic band is isolated. 
To calculate the Zak phase for each type of edge, we use
Eq.~(\ref{zak4}) for the entangled 1st and 2nd photonic bands, Eq.~(\ref{zak2})
for the isolated 3rd photonic band, respectively.

\begin{figure}[ht]
\centering\includegraphics[width=1.0\textwidth]{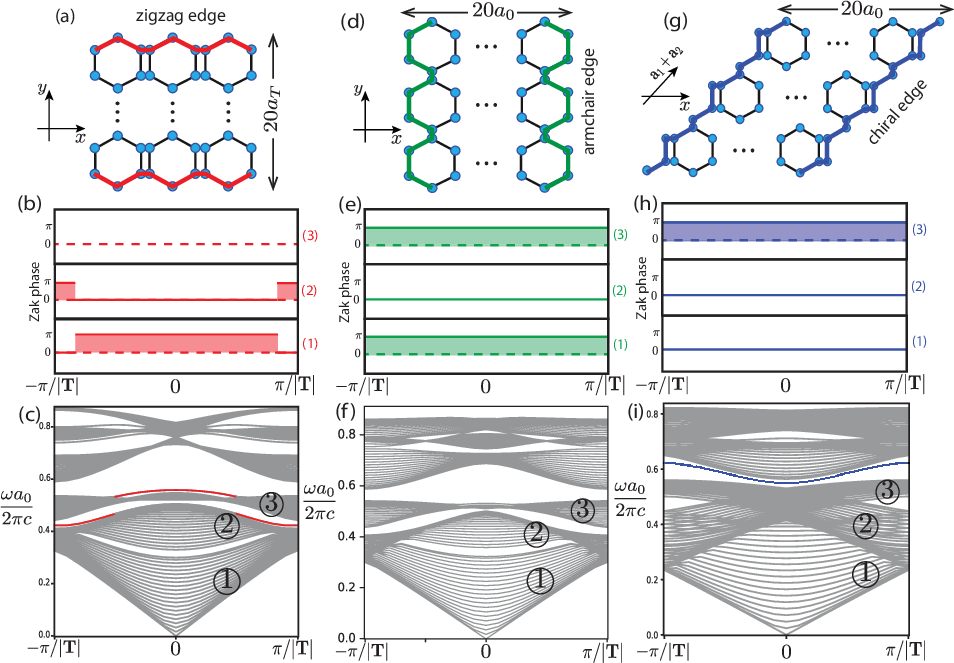}
\caption{(a) 
Schematic structure of a ribbon with zigzag edges, where the $x$-direction is periodic but the $y$-direction is finite with the width $20a_T$.
(b) The Zak phase of the lowest three bands of BPN PhC with zigzag edges.
(c) Photonic band structure for BPN ribbon with zigzag edges.
 The red lines indicate topological edge states. 
(d) Schematic of BPN PhC ribbon with armchair edges, which is periodic
 along $y$-direction and finite in $x$-direction. Its size in $x$-direction is $20a_0$.
(e) The Zak phase of the lowest three bands of BPN PhC ribbon with
 armchair edges. 
(f) Photonic band structure for BPN PhC ribbon with armchair edges. 
(g) Schematic of BPN PhC ribbbon with chiral edges, 
which is periodic along $\mathbf{ T} = \mathbf{a}_1 + \mathbf{a}_2$ direction and finite in
 $x$-direction. This ribbon's width is also $20a_0$ in $x$-direction.
(h) The Zak phase of the lowest three bands of BPN PhC in
 $\mathbf{ T} = \mathbf{a}_1 + \mathbf{a}_2$ direction. 
(i) Photonic band structure for BPN PhC ribbon with chiral edges.
The blue lines indicate topological edge states.}  
\label{figure3}
\end{figure}

Here we shall evaluate the Zak phase of BPN ribbon with zigzag edges and
observe the emergence of topological edge states in photonic band
structure, which are associated with the nonzero Zak phase.
Figure~\ref{figure3}(a) is a schematic of the zigzag ribbon, which is
periodic along the $x$-direction and its width in the $y$-direction is $20a_T$. 
In our numerical calculations, we use the following boundary conditions: 
periodic boundary condition (PBC) for $x$-direction, and  
perfect magnetic conductor (PMC) boundary condition for $y$-direction, respectively.
Figure~\ref{figure3}(b) shows the numerically obtained Zak phase for the
lowest three photonic bands. 
Figure~\ref{figure3}(c) shows the numerically obtained photonic bands
for BPN ribbon with zigzag edges, where the red lines indicate the modes owing
to topological edge states.

As shown in Fig.~\ref{figure3}(b), the 1st and the 2nd photonic
bands have $k$-dependent Zak phase owing to the band inversion at the
degenerate point along the $S-Y$ line in the 1st BZ of bulk 2D BPN. 
The total Zak phase for lowest two bands is $\sum_{n=1}^2 Z^n(k)= \pi$,
which should lead to the topological edge states between the
2nd and the 3rd photonic bands for all $k$ points.
However, since the band gap between the 2nd and the 3rd photonic bands is not a
complete band gap in the present system, the edge states near $k=0$ 
are mixed with bulk states that cannot be detected by numerical calculation.  
Similarly, since the total Zak phase for the lowest three bands is $\sum_{n=1}^3
Z^n(k)= \pi$, the edge states are expected to emerge for all over $k$
between the 3rd and the 4th photonic bands. 
However, it emerges only at the center of the 1st BZ
because of the PMC boundary condition. To realize the topological edge
states in the whole 1st BZ, we need to change the boundary
condition by creating the interface between the zigzag edge and a
trivial PhC, which will be shown later.

Next we shall discuss the case for BPN ribbon with armchair edges.
Figure~\ref{figure3}(d) is the schematic of BPN ribbon with armchair
edges. 
Since the system is only periodic along $y$-direction and its width in the $x$-direction is $20a_0$, we shall
impose the following boundary conditions: 
PBC for $y$-direction and PMC
boundary condition for $x$-direction, respectively.
Figure~\ref{figure3}(f) is the corresponding photonic band structure for
BPN ribbon with armchair edges.
The numerically obtained Zak phase for BPN ribbon with armchair edges is shown
in Fig.~\ref{figure3}(e).  
The 1st and the 3rd photonic bands lead to nontrivial Zak phase with $\pi$,
however, the 2nd photonic bands lead to Zak phase with $0$.
Thus, the total Zak phase for the lowest three bands is $\sum_{n=1}^3 Z^n(k)=2\pi$, which is
equivalent to $0$. 
Therefore, the complete band gap between the 3rd and the 4th photonic band becomes topologically trivial.
As can be seen in Fig.~\ref{figure3}(f), there is no topological edge
state in the complete band gap between the 3rd and the 4th photonic bands, which is consistent with the Zak phase
calculation. Similar to BPN ribbon with zigzag edges, topological edge states
should appear in the photonic band gaps between the 1st and the 2nd bands, and
also between the 2nd and the 3rd bands. 
However, because of the overlapping frequency of these two pairs of
band, we cannot distinguish the topological edge states from bulk states
by numerical calculation.

Finally, we shall consider the case for BPN ribbon with chiral edges
shown in Fig.~\ref{figure3}(g). 
As for BPN ribbon with chiral edges, translational invariance is
preserved along the direction of
$\mathbf{T}=\mathbf{a}_1+\mathbf{a}_2$ and its width in $x$-direction is $20a_0$.
Figure~\ref{figure3}(i) shows the corresponding photonic band structure
for BPN ribbon with chiral edges, in which
the blue lines indicate the modes owing to topological edge states.
In numerical calculation, we shall take the following boundary
conditions: PBC for the translational invariant direction and PMC
boundary condition for the transverse direction, respectively.
Figure~\ref{figure3}(h) shows the numerically obtained Zak phase for the
lowest three photonic bands of BPN ribbon with chiral edges.
Although the Zak phase for the lowest two photonic bands are $0$, it is
$\pi$ for the 3rd band. The complete band gap becomes topologically
nontrivial because the total Zak phase of the lowest three photonic
bands is $\sum_{n=1}^3 Z^n(k)=\pi$. 
Since the photonic band gap between the 3rd and the 4th photonic bands are
topological, the topological edge states emerge as shown in Fig.~\ref{figure3}(i).

\begin{figure}[ht]
\centering\includegraphics[width=0.8\textwidth]{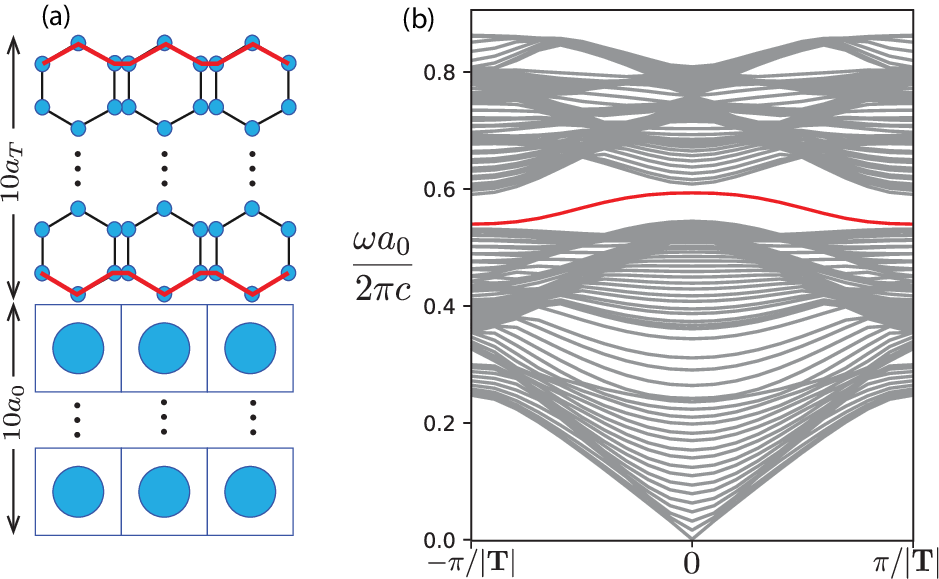}
\caption{(a) Schematic of a structure including 10 BPN PhC unit cells and 10 trivial PhC unit cells in $y$-direction. This structure is periodic in both $x$- and $y$-directions and contains the interface between
BPN ribbon with zigzag edges and the trivial PhC. 
 (b) The corresponding photonic band structure for the
structure of (a). The red lines denote topological edge states, where EM
 wave is highly localized at the interface between the zigzag edge and the
 trivial PhC.}  
\label{figure4}
\end{figure}

To confirm the existence of topological zigzag edge states at all $k$ points, we create another structure containing BPN PhC and a trivial PhC. The trivial PhC has a band gap at the same frequency range as the band gap of BPN PhC. Figure~\ref{figure4}(a) is the schematic of a supercell containing 10 unit cells of BPN PhC and 10 unit cells of trivial PhC in the $y$-direction. This structure has the interface between the zigzag edge and the trivial PhC. Applying the PBC in both $x$- and $y$-directions, we calculate and obtain the photonic band structure for the supercell as shown in Fig.~\ref{figure4}(b). The red lines denote topological edge states, where the EM wave is localized at the interfaces. These edge states emerge at all $k$ points in the 1st BZ and doubly degenerate because the supercell contains two interfaces. 

We have examined three different types of ribbon structures of BPN PhC.
At the complete band gap between bands 3 and 4, the armchair ribbon is topologically trivial, the zigzag and chiral ribbons are topologically nontrivial resulting in the emergence of topological edge states. All the topological edge states are doubly degenerate because each supercell has two equivalent boundaries.

\section{Topological Corner States}\label{sec4}
The topological corner states, as a manifestation of higher-order
topology, are examined. These states involve the confinement of 
EM waves at the corner owing
to the non-zero product of the Zak phase in two edge directions forming
the corner structure~\cite{Liu2021,Liu2019,Ota2019}. In this section, we
examine the localization of EM waves at different corner structures in the BPN
PhC lattice. 

\begin{figure}[ht!]
\centering\includegraphics[width=0.9\textwidth]{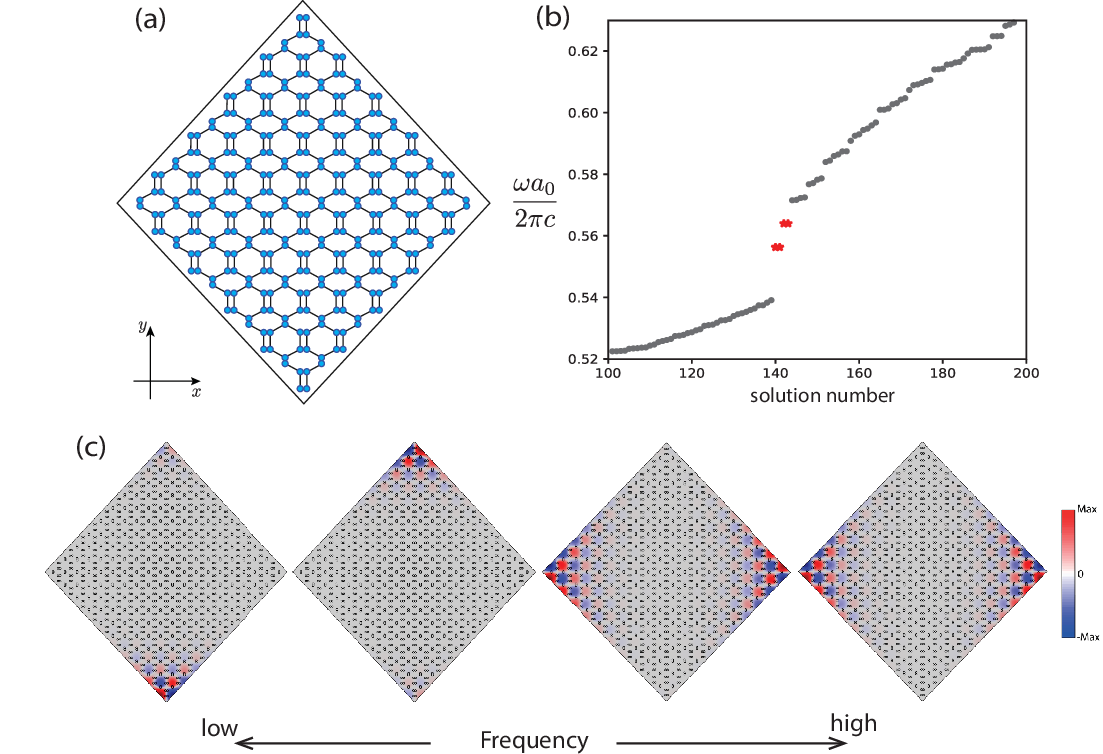}
\caption{(a) Schematic of corner structure which is formed by chiral
edges (type I). (b) Frequency spectrum for type I. 
The red stars indicate topological corner states. The frequency
 range of corner states is in the complete gap region. (c) The field
 profile for four corner states from low to high frequencies. These
 profiles show that topological corner states is isolated from bulk and
 edge states.} 
\label{figure5}
\end{figure}

\begin{figure}[ht!]
\centering\includegraphics[width=0.7\textwidth]{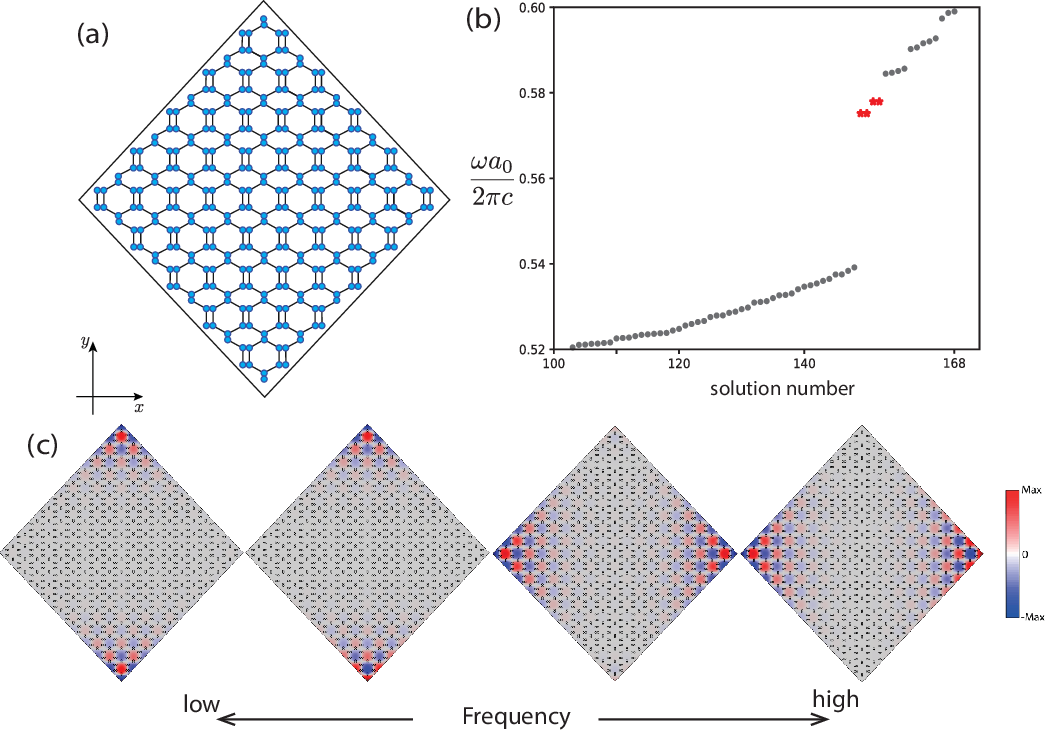}
\caption{(a) Schematic of corner structure of type II.
Edges are chiral edges same as type I, however, the detail structures
 around corner are different.
 (b) Frequency spectrum for type II.
 The red stars indicate topological corner states. The frequency
 range of corner states is in the complete gap region. (c) The field
 profile for four corner states from low to high frequencies. These
 profiles show that topological corner states is isolated from bulk and
 edge states.} 
\label{figure6}
\end{figure}

In Fig.~\ref{figure5}(a), we show a rhombus-shaped BPN PhC. Four outer
boundaries have the chiral edges. 
Since there are two possible ways to construct a rhombus-shaped BPN PhC with chiral edges,
to distinguish this rhombus
structure with the latter one, we refer to as "type I". 
The two opposite corners have the same structure and they are different from two other
ones. The PMC boundary condition is applied for all edges in this
rhombus structure.
Figure~\ref{figure5}(b) is the frequency spectrum for the type I rhombus-shaped BPN PhC. 
In the complete band gap of BPN PhC, there are
four isolated states labeled by red stars. 
These states are corner states where the EM wave is highly localized at
the corners and exponentially decays as shown in Fig.~\ref{figure5}(c).  
Since the corner structure are formed by the two chiral edges which are
topological, the product of Zak phases among these two chiral edges gives
nonzero Zak phase $\pi$. Thus, the corner state are topologically
protected states, because the topological corner states based on the
same condition are observed in other topological systems~\cite{Ota2019, Liu2021}.

Similarly, we can construct another corner structure from two chiral
edges, which we refer to as "type II" as shown in
Fig.~\ref{figure6}(a). 
The difference of type II structure from type I can be seen 
at the corner sites. 
Figure~\ref{figure6}(b) shows the frequency spectrum for type II,
where four isolated states labeled by red stars are numerically observed
in the complete band gap. 
These states are also topological corner states according to the same
reason of type I. 
%
Their EM waves distribute mainly at the corners, and decay
exponentially as shown in Fig.~\ref{figure6}(c), in which
the frequency increases from left to right panels. 

\begin{figure}[ht]
\centering\includegraphics[width=0.9 \textwidth]{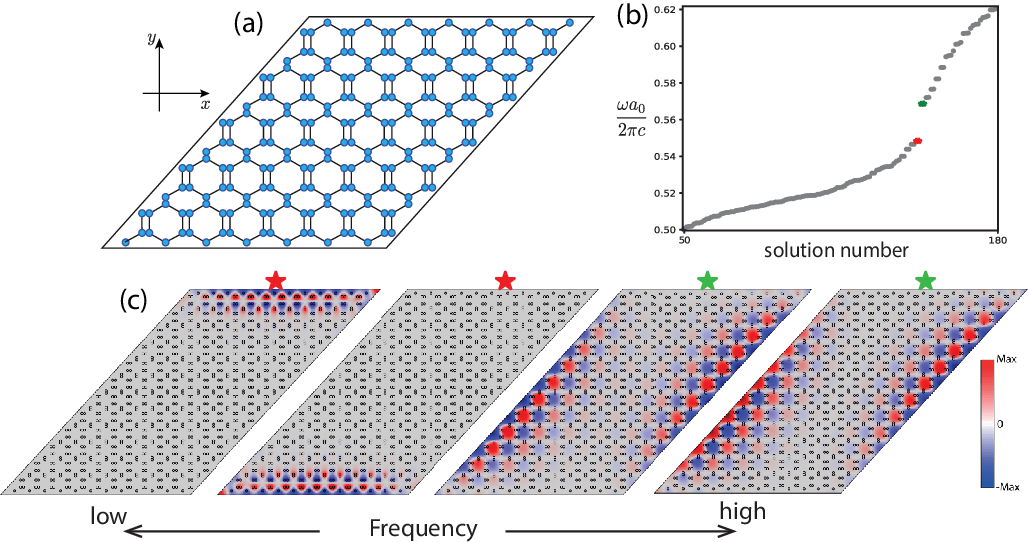}
\caption{(a) Schematic of corner structure formed by zigzag and chiral edges. (b) Photonic band structure for the corner structure formed by
 the zigzag and the chiral edges. The red stars indicate the zigzag
 edge states, the green stars denote the edge states of chiral edge. (c)
 The field distribution for the states labeled in red and 
 green from low to high frequencies.} 
\label{figure7}
\end{figure}

Finally, we examine another corner structure which is formed by the
zigzag and the chiral edges as shown in Fig.~\ref{figure7}(a). 
The left and
right are chiral edges, the upper and lower are zigzag edges. Because
both the zigzag and 
chiral edges are topologically nontrivial at the complete band gap, the
topological corner states are expected to emerge in this
structure. Figure~\ref{figure7}(b) displays photonic band structure for the
corner structure containing zigzag and chiral edges. 
The states below the band gap (labeled with red stars) represent the
edge states at the zigzag edges, while the states above the band gap
(labeled with green stars) correspond to the edge states at the chiral
edges.
Their field distributions are shown in 
Fig.~\ref{figure7}(c),
where the frequency increases from left to right panels. 
This frequency range is consistent with the frequency range of the edge
states for the BPN edges with zigzag edges and BPN edges with chiral edges,  
as shown in Fig.~\ref{figure3}(f) and (i), respectively. 
However, no corner states are found in this structure.  

When we refer to existing research on 2D crystal structures where
topological corner states emerge~\cite{Ota2019, Phan2021,
Liu2021,Li2020}, we find that the two edges forming the 
corner are topologically nontrivial, and their frequency levels are
identical. This allows them to interact with each other to form
topological corner states. The reason for the absence of topological
states in the present structure is that the frequency ranges of the
zigzag and chiral edge states are entirely different, preventing any
interaction between the two edge states. Therefore, electromagnetic
waves cannot be confined to the boundary between two edges.

Other corner structures involving the armchair edge, such as corners
between the zigzag and armchair edges or corners between the armchair
and the chiral edges, can also be constructed. However, topological
corner states do not exist in these structures due to the trivial
properties within the complete band gap of the armchair edge.

\section{Summary}\label{sec5}
In this paper, we have numerically studied EM waves in BPN PhC structure
by using finite difference and finite element methods. The original
photonic analog of the BPN has no band gap. Linear dispersion of
frequency can be observed in this PhC. When the structure is modified by
increasing the distance from each rod to the center of the unit cell, a
complete band gap opens in between the 3rd and the 4th photonic bands. 
We examine the topological properties of BPN PhC in this photonic band
gap by numerically calculating the Zak phase for several different edge
structures. 
The topological edge states are observed in the photonic band gap due to
the nontrivial Zak phase, which are localized at the zigzag
and the chiral edges.  
The higher-order topology as the corner states are found at the corner
formed by two chiral edges because of non-zero product of the Zak phase
in two directions. We also point out that the topological corner states
emerge only when the topological edge states of two edge structures
which form the corner are completely overlapped to each other. 

Compared with graphene-like PhCs, the topological edge states in
BPN-like structures can be found without breaking the symmetry of the
crystals. Our results suggest a possible way to design in-gap
topological waveguide and topological confinement of EM waves.

\begin{backmatter}
\bmsection{Funding}
Japan Society for the Promotion of Science (22H05473, 21H01019, JP18H01154); Japan Science and Technology Agency (JPMJCR19T1).

\bmsection{Acknowledgments}
K.W. acknowledges the financial support by JSPS KAKENHI (Grant No. 22H05473, 21H01019 and JP18H01154), and JST CREST (Grant No. JPMJCR19T1). 

\bmsection{Disclosures}
The authors declare no conflicts of interest.

\bmsection{Data Availability Statement}
Data underlying the results presented in this paper are not publicly available at this time but may
be obtained from the authors upon reasonable request.

\end{backmatter}


\bibliography{reference}






\end{document}